\documentclass[a4paper,fleqn]{cas-dc}
\usepackage[numbers]{natbib}

\usepackage{pgfplots}
\usepackage{pgfplotstable}
\usepackage{tikz}
\usepackage{subcaption}
\usepackage[flushleft]{threeparttable}
\usepackage{graphicx}
\usepackage{amsmath}
\usepackage{color}
\pgfplotsset{width=7cm,compat=1.15}
\def\tsc#1{\csdef{#1}{\textsc{\lowercase{#1}}\xspace}}
\tsc{WGM}
\tsc{QE}
\tsc{EP}
\tsc{PMS}
\tsc{BEC}
\tsc{DE}

\begin{document}
\let\WriteBookmarks\relax
\def\floatpagepagefraction{1}
\def\textpagefraction{.001}
\shorttitle{Journal of Parallel and Distributed Computing }
\shortauthors{Qiong Chang et~al.}

\title [mode = title]{Multi-directional Sobel operator kernel on GPUs}

\author[1]{Qiong Chang}[orcid=0000-0002-4447-0480]\cormark[1]
\address[1]{School of Computing, Tokyo Institute of Technology, Japan}
\ead{q.chang@c.titech.ac.jp}
\author[2]{Xiang Li}
\address[2]{School of Electronic Science \& Engineering, Nanjing University, China}
\author[2]{Yun Li}\cormark[1]
\ead{yli@nju.edu.cn}
\author[1]{Jun Miyazaki}
\cortext[cocitation8]{Corresponding author}

\begin{abstract}
Sobel is one of the most popular edge detection operators used in image
processing. To date, most users utilize the two-directional $3\times{3}$ Sobel
operator as detectors because of its low computational cost and
reasonable performance.
Simultaneously, many studies have been conducted on using large multi-directional
Sobel operators to satisfy their needs considering the high stability, but at
an expense of speed.
This paper proposes a fast graphics processing unit (GPU) kernel for the four-directional 5x5 Sobel
operator. To improve kernel performance, we implement the kernel based on
warp-level primitives, which can significantly reduce the number of memory accesses.
In addition, we introduce the prefetching mechanism and operator
transformation into the kernel to significantly reduce the computational complexity and
data transmission latency.
Compared with the OpenCV-GPU library, our kernel shows high performances of 6.7x 
speedup on a Jetson AGX Xavier GPU and 13x on a GTX 1650Ti GPU.

\end{abstract}




\maketitle

\section{Introduction}
The Sobel operator is a classical first-order edge detection operator that performs a 2D spatial gradient 
measurement on images and is generally used to find the approximate absolute gradient 
magnitude at each pixel. It is typically used to emphasize regions of high spatial frequency that 
correspond to edges. Compared with other edge detection operators, such as the Canny and Roberts 
cross, the Sobel operator has a low calculation amount, simple structure, and high precision. 
Therefore, it has a wide range of applications in fields such as remote sensing~\cite{citation26}, medical image processing~\cite{citation27}, and industrial detection~\cite{citation1}.
To date, most applications using Sobel have chosen the two-directional
$3\times{3}$ operator as their detectors because of its low computational cost
and reasonable performance.
Nevertheless, some applications still require further size expansions and increases in the direction of the operator to satisfy their unique
requirements.
In the medical field, Sheik et al.~\cite{citation2} proposed a Sobel operator for the
edge detection of the knee-joint space of osteoarthritis. They improved the operator
by adding $315^\circ$ and $360^\circ$ directions on the bias of horizontal and
vertical directions, which can perform the detection better than the original two.
Remya et al.~\cite{citation3} applied the Sobel operator to the edge detection of
brain tumors in MRI images. Their operator was improved to handle eight directions to clearly detect extremely irregularly shapes of
tumors and showed a higher detection accuracy than other methods.
In the industrial field, Min et al.~\cite{citation4} utilized the Sobel operator to
detect the edges of screw threads. Considering the type of screw thread angles are mainly $30^\circ$, $55^\circ$ and $60^\circ$, they assigned spatial weights and added $67.5^\circ$ and
$112.5^\circ$ directions for the operator, which can efficiently extract more precise
edges and achieve better continuity than conventional methods.
In addition to the direction, Siyu et al.~\cite{citation5} expanded the operator size from
$3\times{3}$ to $7\times{7}$ using the average gradient vectors of two
neighboring pixels to quantify aggregate angularity. Compared with conventional methods using one pixel, the improved method helps calculate a more stable angularity index value.
All these applications demonstrated that, in some cases, a large
multi-directional Sobel operator has higher robustness than the traditional
operator and can better adapt to actual requirements. However, as mentioned in
~\cite{citation2}, it always requires more computing time.
In particular, as the image size increases, the amount of computation increases
exponentially, which burdens applications using edge
detection as a preprocessing step.
This paper proposes a fast graphic processing unit (GPU) kernel for a four-directional
$5\times{5}$ Sobel operator, because GPUs usually show an excellent performance on real-time image processing problems~\cite{citation6}~\cite{citation7}. In our experience, a 5$\times$5 Sobel operator has similar edge detection robustness to a 7$\times$7 operator but higher than 3$\times$3. Meanwhile, the processing speed of a multi-directional 5$\times$5 operator is significantly slower than a 3$\times$3 operator~\cite{citation8}.
To improve kernel performance, we made innovations in the following aspects:
\begin{itemize}
\item we implement the kernel based on warp-level primitives, which can significantly
  reduce the number of memory accesses;
\item we provide an efficient procedure with the prefetching mechanism,
  which significantly reduces the computational complexity and data transmission latency, and
\item we further accelerate the operations in diagonal directions using a two-step optimization approach, 
which helps to increase the data reuse rate.
\end{itemize}
The remainder of this paper is organized as follows. Section 2 introduces the
acceleration strategies and results of the current Sobel operator. Section 3
provides the principle of the four-directional $5\times{5}$ Sobel operator. In Section 4, 
we introduce the implementation and optimization details of our GPU kernel. Then, we evaluate the kernel performance 
in Section 5 and finally conclude this paper in Section 6.

\section{Related Work}
Recently, several studies have been conducted on accelerating the Sobel
edge detection using GPUs.
Jo et al.~\cite{citation9} optimized their GPU-based Sobel kernel using shared
memory. They assigned the entire image to several blocks and used the
corresponding streaming multiprocessors (SMs) to detect edges in parallel with
their respective local information. This approach is simple and versatile but
has limited performance improvement because the overuse of shared memory
typically affects the number of active blocks and reduces the parallelism of the
kernel. Nevertheless, the shared memory approach is 1.65x faster than that of global memory.
Chouchene et al.~\cite{citation10} realized a fast grayscale and Sobel edge detection on
GPUs, which was approximately 50x faster than running on a CPU. Similar to~\cite{citation9},
they split the edge detection task to each SM and stored the image fragments in
the corresponding shared memory. The difference is that they enabled different
sizes of CUDA blocks to accomplish the detection task, which proved more
efficient in their application than the block-size consistent method.
Xiao et al.~\cite{citation11} proposed an eight-directional Sobel operator on GPU using
the open computing language (OpenCL) framework. In their implementation, each
work item handles the convolution calculations for four pixels instead of one,
which can significantly improve memory access efficiency and reduce computational complexity. Compared with approaches implemented by CPU,
OpenMP, and CUDA, they are 9.55, 2.23, and 1.17x faster, respectively.
Zuo. et al.~\cite{citation12} implemented a fast Sobel operator on GPUs. They optimized their GPU kernel as follows: 1) using the texture memory to
store image data to accelerate memory access; 2) performing a single
thread to process the calculations of multiple pixels, which can significantly
increase the overall throughput and 3) fully exploiting the symmetry of
Sobel operators to reuse intermediate results to reduce the entire computational
complexity. They achieved a 122x acceleration ratio for a $4096\times{4096}$
image compared with the CPU-based implementation. However, owing to the limitation of texture memory size, it is unreasonable for common situations that require multiple images in
practical applications.
The purposes of all the these methods are to accelerate the Sobel-based edge
detection as fast as possible, while ensuring the correctness of results to 
satisfy real-time processing requirements.
However, owing to the limitations of their parallel algorithms under GPU
architectures, much room remains for improving their acceleration
methods. Then, we will introduce an in-depth acceleration method for the Sobel operator.
\begin{figure}
\centering
\includegraphics[width=3in]{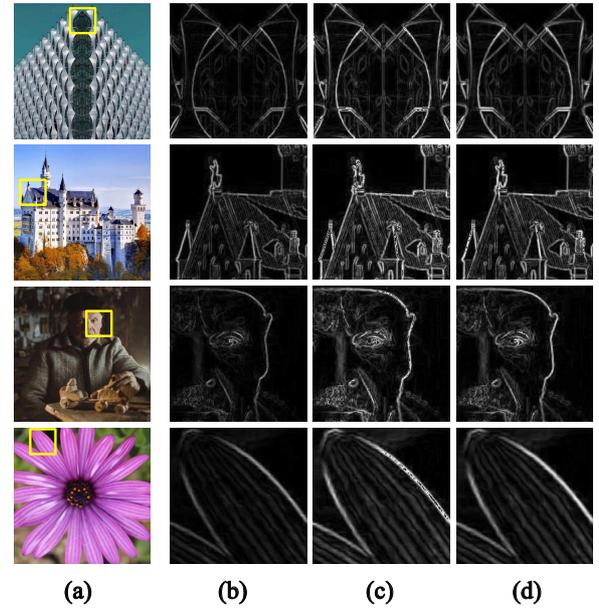}
\caption{Edge detection results. (a) Original image. (b) Two-directional
  $3\times{3}$. (c) Four-directional $3\times{3}$. (d) Four-directional
  $5\times{5}$.}
\label{fig:edge_detection}
\end{figure}
\section{Four-Directional $5\times{5}$ Sobel Operator}
\subsection{Operator Definition}
The Sobel operator is a classical edge detection operator proposed by Irwin Sobel
and Gary Feldman~\cite{citation13}.
It is a discrete differential operator that detects the edge features of images
by computing the pixel gradients.
The original Sobel operator is an isotropic gradient operator using two $3\times{3}$
filters to convolve with an image and obtain derivative approximations: one each for
horizontal and vertical changes.
Equation~\ref{equ:sobel3_1} shows the basic computation of the Sobel operator:
\begin{equation}
  G_x=\begin{bmatrix} -1&\textbf{0}&1\\-2&\textbf{0}&2\\-1&\textbf{0}&1\end{bmatrix}\ast{I}\\,
  G_y=\begin{bmatrix} -1&-2&-1\\\textbf{0}&\textbf{0}&\textbf{0}\\1&2&1\end{bmatrix}\ast{I},
\label{equ:sobel3_1}
\end{equation}
where $I$ represents the input image, and $G_x$ and $G_y$ are the two images
containing the horizontal and vertical derivative approximations, respectively.
$\ast$ denotes the basic convolution calculation between the input
image and the two filters.
In these two filters, because the weights of the central axis in both directions
are $0$, and the two sides in both directions are opposite to each other, the
convolution results are equivalent to calculating the differences between the
two sides, which means calculating the gradients in both directions.
Then, the final result can be aggregated by calculating the root sum of square
(RSS) of $G_x$ and $G_y$ as follows:
\begin{equation}
 G=\sqrt{G_x^2+G_y^2}.
\label{equ:sobel3_2}
\end{equation}

Figure~\ref{fig:edge_detection}(a) shows the original images, and
Fig.~\ref{fig:edge_detection}(b) shows the local edge images (the yellow boxes in
Fig.~\ref{fig:edge_detection}(a)) detected using the original two-directional
$3\times{3}$ Sobel operator. Although some texture information is lost, the
overall contour of the petals, houses, and figures are clearly preserved.

The original Sobel operator considers only the horizontal ($0^\circ$) and vertical ($90^\circ$) directions. To further enhance its effect, we introduce a
$5\times{5}$ Sobel operator by adding two diagonal directions ($45^\circ$ and
$135^\circ$). The filters of the four-directional $5\times{5}$ Sobel
operator can be defined as follows:
\begin{equation}
\begin{aligned}
  G_x&=\begin{bmatrix} -1&-2&\textbf{0}&2&1\\-4&-8&\textbf{0}&8&4\\-6&-12 &\textbf{0} &12 &6\\-4&-8&\textbf{0}&8&4\\-1&-2&\textbf{0}&2&1\end{bmatrix}\ast{I},\\
  G_y&=\begin{bmatrix} -1&-4&-6&-4&-1\\-2&-8&-12&-8&-2\\\textbf{0}&\textbf{0} &\textbf{0} &\textbf{0} &\textbf{0}\\2&8&12&8&2\\1&4&6&4&1\end{bmatrix}\ast{I},\\
  G_d&=\begin{bmatrix} -6&-4&-1&-2&\textbf{0}\\-4&-12&-8&\textbf{0}&2\\-1&-8 &\textbf{0} &8 &1\\-2&\textbf{0}&8&12&4\\\textbf{0}&2&1&4&6\end{bmatrix}\ast{I},\\
  G_{dt}&=\begin{bmatrix} \textbf{0}&-2&-1&-4&-6\\2&\textbf{0}&-8&-12&-4\\1&8 &\textbf{0} &-8 &-1\\4&12&8&\textbf{0}&-2\\6&4&1&2&\textbf{0}\end{bmatrix}\ast{I},
\end{aligned}
\label{equ:sobel5_1}
\end{equation}
and the final results can be aggregated as follows:
\begin{equation}
 G=\sqrt{G_x^2+G_y^2+G_d^2+G_{dt}^2},
\label{equ:sobel5_2}
\end{equation}
where $G_d$ and $G_{dt}$ represent the two images containing the diagonal
derivative approximations. They can be obtained by rotating $G_x$ and $G_y$
by $45^\circ$. In general, users can define the filter weights 
according to their needs and must only combine the Gaussian smoothing and differentiation.
In Eq.~\ref{equ:sobel5_1}, the weight values are generated using the OpenCV Sobel library and used to perform the edge detection shown in
Fig.~\ref{fig:edge_detection}(d). To better distinguish the effect from
the $3\times{3}$ operator, the detection results obtained by the four-directional
$3\times{3}$ operator are listed in Fig.~\ref{fig:edge_detection}(c).
Compared with the two-directional operator, the four-directional operator
provides more abundant textures, such as petals and buildings. Furthermore, $5\times{5}$
operator is more insensitive to surrounding changes and less affected by
noise, which helps provide clearer edge features and higher robustness than using
$3\times{3}$.
However, a four-directional $5\times{5}$ operator without optimization is approximately eight times more computationally
intensive than the two-directional $3\times{3}$, which significantly slows the processing speed. Therefore, developing an efficient
acceleration method for the four-directional $5\times{5}$ operator is essential.
\subsection{Filter Weight Generalization}
To avoid limiting our method to the constant weight values in
Eq.~\ref{equ:sobel5_1}, we generalize our Sobel operator as follows:
\begin{equation}
\begin{aligned}
  K_x&=a\cdot{\begin{bmatrix} 1\\n\\m\\n\\1\end{bmatrix}}\times{\begin{bmatrix}-1&-b&\textbf{0}&b&1\end{bmatrix}}\\
  &=a\cdot{\begin{bmatrix} -1&-b&\textbf{0}&b&1\\-n&-nb&\textbf{0}&nb&n\\-m&-mb &\textbf{0} &mb &m\\-n&-nb&\textbf{0}&nb&n\\-1&-b&\textbf{0}&b&1\end{bmatrix}}=(k_{ij})_{5\times5},\\
  K_y&=a\cdot{\begin{bmatrix} -1\\-b\\\textbf{0}\\b\\1\end{bmatrix}}\times{\begin{bmatrix}1&n&m&n&1\end{bmatrix}}\\
  &=a\cdot{\begin{bmatrix} -1&-n&-m&-n&-1\\-b&-nb&-mb&-nb&-b\\\textbf{0}&\textbf{0} &\textbf{0} &\textbf{0} &\textbf{0}\\b&nb&mb&nb&b\\1&n&m&n&1\end{bmatrix}}=(k_{ij})_{5\times5},\\
  K_d&=a\cdot{\begin{bmatrix} -m&-n&-1&-b&\textbf{0}\\-n&-mb&-nb&\textbf{0}&b\\-1&-nb &\textbf{0} &nb &1\\-b&\textbf{0}&nb&mb&n\\\textbf{0}&b&1&n&m\end{bmatrix}}=(k_{ij})_{5\times5},\\
  K_{dt}&=a\cdot{\begin{bmatrix} \textbf{0}&-b&-1&-n&-m\\b&\textbf{0}&-nb&-mb&-n\\1&nb &\textbf{0} &-nb &-1\\n&mb&nb&\textbf{0}&-b\\m&n&1&b&\textbf{0}\end{bmatrix}}=(k_{ij})_{5\times5},\\
  &a\in{\mathbb{Z^+}},\hspace*{0.5cm} b,m,n\in{\mathbb{R^+}},\hspace*{0.5cm} and\hspace*{0.5cm}   \forall{k_{ij}}\in{\mathbb{Z}}.
\end{aligned}
\label{equ:sobel5_param}
\end{equation}
In these filters, {\em a}, {\em b}, {\em m}, and {\em n} are all positive
numbers, and all the items $k_{ij}$ are integers. The generalized operator ensures that
the absolute values of the weights remain symmetrical in the horizontal and
vertical directions, and the positive and negative relationship is
unchanged. Here, a constraint is added to the weight values: expressing $K_x$ and
$K_y$ as a constant {\em a} multiplied by two vectors containing
1, which means that the filter weight values are proportional in both
horizontal and vertical directions. This constraint is expected to help us reuse
the intermediate results and improve computational efficiency without affecting
the Sobel edge detection. $K_d$ and $K_{dt}$ do not satisfy this rule,
requiring further optimization in Section~\ref{sec:d_optimization}.

\section{GPU Implementation}
In this section, we introduce our acceleration strategies from three
aspects: 1) the kernel implementation method based on warp-level primitives,
2) the procedure for the entire image using the prefetching mechanism, and 3)
the operator transformation for diagonal directions.
However, we first introduce two key GPU techniques used in our
optimization method.
\subsection{GPU Warp-level Primitives}
Nvidia GPUs and the CUDA programming model use the single instruction,
multiple threads (SIMT) execution model to maximize the computing capability of
GPUs~\cite{citation14}. GPUs execute warps of 32 parallel threads using
SIMT, enabling each thread to access its registers, load and store
from divergent addresses, and follow divergent control flow paths. In addition, the CUDA
compiler and GPUs work together to ensure the threads of a warp execute the identical
instruction sequences together as fast as possible.
Current CUDA programs can achieve high performances using explicit
warp-level primitives, such as warp shuffles. Warp shuffles are a fast
mechanism for shifting and exchanging register data between threads in the
same warp, such as $\_\_shfl\_down\_sync$ and $\_\_shfl\_xor\_sync$. These instructions 
can efficiently complete the reduction and scan operations for the data stored in 
vector registers without using other types of memory.
Using the prefetching mechanism can save data latency and increase
practical thread utilization, significantly improving program 
performance. 
This study fully uses this mechanism to implement our GPU kernel.

\subsection{Prefetching Mechanism}
\label{sec:prefetching}
The prefetching mechanism is a standard technology used in CPUs to hide the latency of memory operation. 
The processor caches data and instruction blocks before they are executed. While that data travels to the execution units, other instructions can be executed simultaneously.
GPUs also support the prefetching mechanism, which has higher costs than CPUs. 
Although GPUs typically use excess threads to hide memory latency, 
using the prefetching mechanism is an excellent decision through 
explicit instructions, which require frequent access 
to the global memory and loading part of data each time~\cite{citation15}.

\subsection{GPU Kernel Design}
We now introduce the details of our GPU kernel. The preparations 
for the input image, including grayscale, boundary padding, and transmission, are treated the same as in~\cite{citation16}.
\subsubsection{Task Assignment}
As shown in Fig.~\ref{fig:sobel_kernel}(a), the input image is evenly
distributed to different blocks for parallel processing by the GPU. Each 
block is assigned multiple rows and columns of image data. Because 
the Sobel filter is a surrounding window centered on the target pixel, 
two adjacent blocks must have overlaps. When the radius of the filter 
is $r$, the overlap between any two blocks is $2r$.

\subsubsection{Data Flow}
For a large input image, assigning a thread to each 
pixel is expensive. Our strategy involves allocating sufficient threads in the horizontal
direction while processing sequentially in the vertical direction. Moreover,
because the Sobel operator does not require frequent data sharing between pixels, 
the shared memory is not used in our kernel. This avoids reducing 
block parallelism caused by excessive allocation of shared memory and reduces 
latency caused by memory accesses.
Figure~\ref{fig:sobel_kernel}(b) shows the data flow of one block for filter 
$K_x$. At the beginning, {\em 2r+1} rows (5 rows for a $5\times{5}$ operator) of 
the input image are loaded into the kernel sequentially and processed to achieve the 
detection result of the {\em r}th row (ROW 0). Then, for each incremental row 
(ROW 1,2) in the output image, only the incremental parts of the input image (row 5,6) 
must be updated at a time. This is primarily because the filter $K_x$ can be decomposed 
into the product of the two vectors shown in Eq.~\ref{equ:sobel5_param}, indicating that 
the calculations of the horizontal and vertical directions can be performed separately. 
Therefore, the intermediate results of the overlapping rows (rows 2 - 4) can be held 
in the kernel, and only incremental rows must be calculated each time.
As mentioned, because each block has overlapping regions, the output 
image size is smaller than the input, with {\em 2r} fewer columns and rows in the 
horizontal and vertical directions, respectively.

\subsubsection{Process Detail}
\label{sec:process_detail}
\begin{figure*}
\centering \includegraphics[width=6.6in]{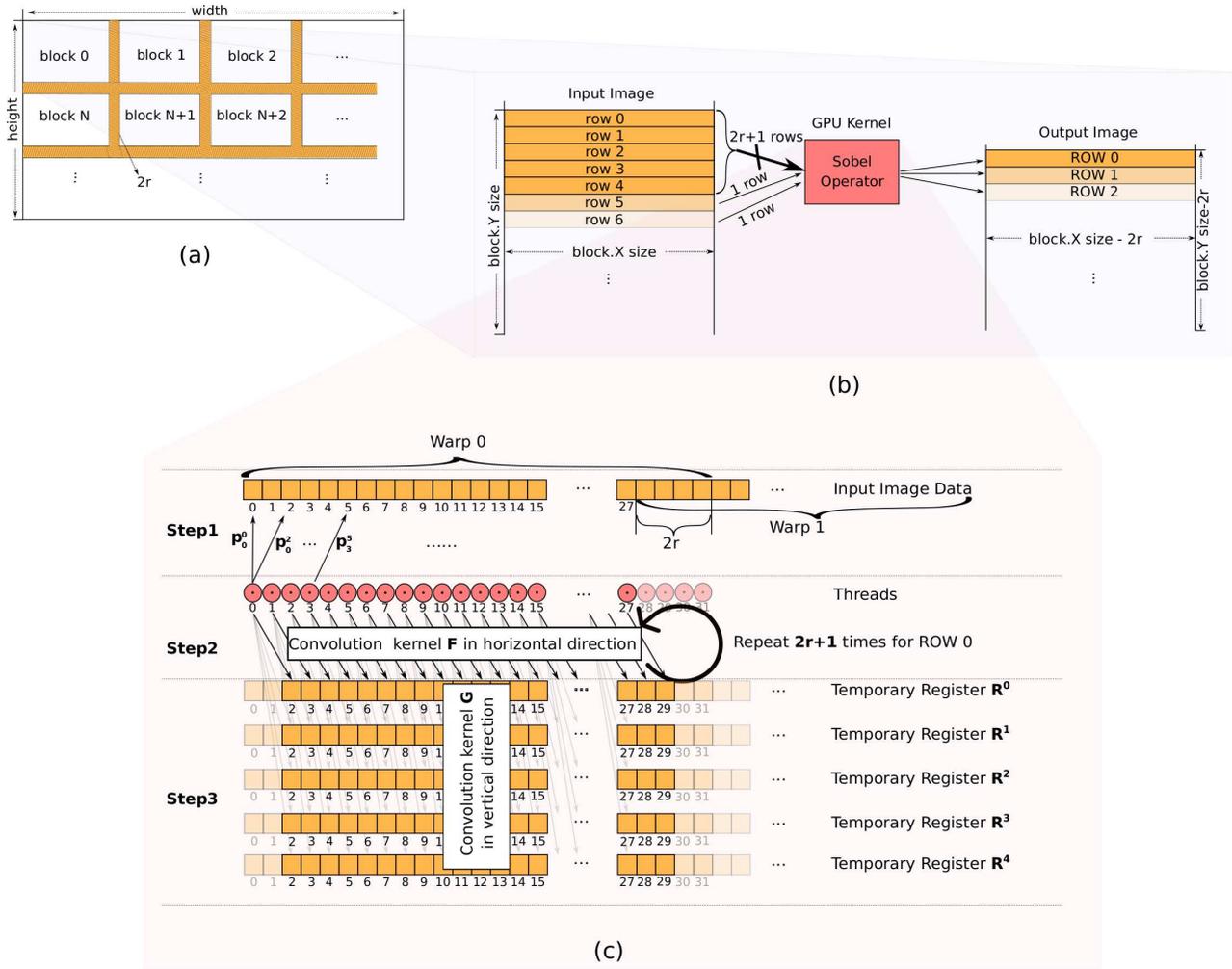}
\caption{Implementation of 5$\times$5 Sobel filter $K_x$. (a) Task
  assignment. (b) Data flow. (c) Process detail.}
\label{fig:sobel_kernel}
\end{figure*}

Figure~\ref{fig:sobel_kernel}(c) shows the process detail of one warp for filter $K_x$. 
The actions can be divided into three steps as follows.
\begin{itemize} 
\item Step 1: each thread loads the corresponding pixel data in one row from the 
global memory to the register, and then shares it with other threads within the 
same warp using the $\_\_shfl\_down\_sync$ primitive. Here, we define the $p_i^j$ 
to denote the obtained pixel data, where {\em i} denotes the thread ID and 
{\em j} denotes the pixel index. Because for a $5\times{5}$ filter, the upper 
bound of {\em j} is {\em i+4}, and data sharing between threads cannot cross the 
warp, the last four ({\em 2r}) threads will be idle, which is the reason the 
overlap between blocks is {\em 2r} columns.
\item Step 2: after obtaining the necessary pixel data, each thread performs the
basic convolution operations as follows:
  \begin{equation}
    F_i^u=-1\cdot{p_i^0}+(-b)\cdot{p_i^1}+b\cdot{p_i^3}+p_i^4,
    \label{equ:Fx}
  \end{equation}
  and stores result $F_i^u$ to the corresponding register $R_i^u$, where {\em u} 
  denotes the row index of the input image. Then, for the initial calculation, 
  Steps 1 and 2 are repeated for {\em 2r+1} times until all rows are calculated. 
  Otherwise, only the oldest register data needs to be updated. Note that because our method is based on separable convolution, 
  instead of expanding the calculation around a target pixel, thread {\em i} 
  actually calculates the result of pixel {\em i+r} in each row. 
\item Step 3: after completing horizontal calculations, each thread begins 
to perform the vertical convolution operations $G_i$, whose equation can be expressed 
as follows:
  \begin{equation}
    \begin{aligned}
      G_i^v&=a\cdot{F_i^{f(v-2)}}+an\cdot{F_i^{f(v-1)}}+am\cdot{F_i^{f(v)}}\\
      &+an\cdot{F_i^{f(v+1)}}+a\cdot{F_i^{f(v+2)}},\\
    \end{aligned}
    \label{equ:Gx}
  \end{equation}
where
  \begin{equation}
    f(x)=x\bmod{5}, \hspace*{0.5cm} x\geq{0}.
    \label{equ:Fx_index}
  \end{equation}
\end{itemize}
$f(x)$ is used to represent the row index because $F$ obtained by the
Eq.~\ref{equ:Fx} is dynamically updated. The $v$ in Eq.~\ref{equ:Gx} denotes 
the index of the center row, always maintaining the variable $x$ greater than 0.
Referring to the three steps, edge detection in the horizontal
direction can be effectively implemented using the $5\times{5}$ Sobel filter 
$K_x$. Similarly, detection using filter $K_y$ in the vertical direction can be 
implemented in the same manner, using different coefficients.

\subsubsection{Optimization for Data Loading}
\begin{figure}
\centering \includegraphics[width=3.3in]{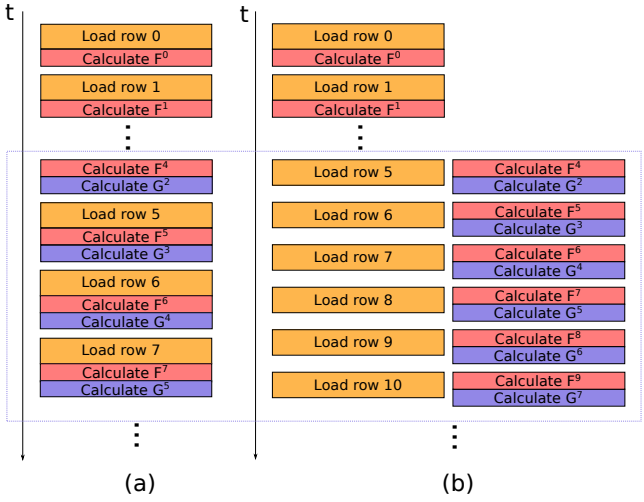}
\caption{Optimization for data loading. (a) Sequential execution. (b) Prefetching.}
\label{fig:prefetching}
\end{figure}

Typically, the data loading of the increments and calculations can be sequentially alternated as shown in Fig.~\ref{fig:prefetching}(a). Its benefit is in avoiding occupying too many on-chip registers, thereby reducing the thread parallelism.
However, it also directly leads to frequent accesses to the global memory, which 
can generate a considerable latency.
To solve this, we explicitly load the incremental image data from the 
global memory while calculating the on-chip data using the prefetching mechanism 
mentioned in Section~\ref{sec:prefetching}.
As shown in Fig.~\ref{fig:prefetching}(b), after loading the fourth row, we 
continue with loading the fifth row without directly completing the calculations of 
{\em G}. Instead, these calculations will end while waiting for the loading 
to complete, which can help us achieve parallelism in time and significantly improve 
the efficiency of kernel execution.
Here, because the prefetching trades more registers for time parallelism, 
to avoid additional burden to the processor, only one row of image data is 
fetched at a time. Thus, the row index function {\em f(x)} in
Eq.~\ref{equ:Fx_index} is changed to
\begin{equation}
   f(x)=x\bmod{6}, \hspace*{0.5cm} x\geq{0},
\label{equ:Fx_index_new}
\end{equation} 
when the prefetching mechanism is active.

\subsubsection{Optimization for Diagonal Direction}
\label{sec:d_optimization}
\begin{figure*}
\centering \includegraphics[width=6.7in]{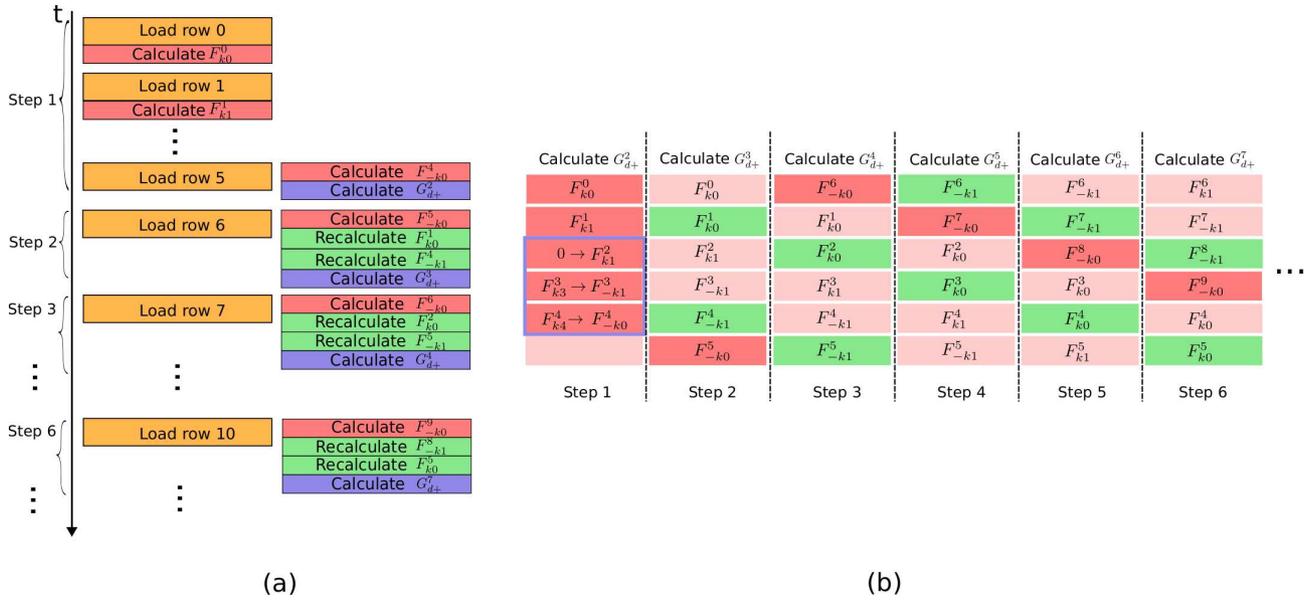}
\caption{Calculation for $G_{d+}$. (a) procedure. (b) Registers status.}
\label{fig:kd+}
\end{figure*}

According to Eq.~\ref{equ:sobel5_param}, elements $k_{ij}$ of $K_d$ and
$K_{dt}$ are neither symmetric nor proportional, which implies that they cannot be
directly decomposed in the same manner as $K_x$ and $K_y$. This also
implies that the convolution results $F_d$ and $F_{dt}$ in horizontal cannot be
reused and must be recalculated for each $G_d$ and $G_{dt}$.
To solve this, we propose a new idea to generate two matrices $K_{d+}$
and $K_{d-}$ as follows.
\begin{equation}
\begin{aligned}
  K_{d+}&=K_d+K_{dt}\\ 
  &=a\cdot{\begin{bmatrix}-m&-n-b&-2&-n-b&-m\\
      b-n&-mb&-2nb&-mb&b-n\\
      \textbf{0}&\textbf{0}&\textbf{0} &\textbf{0}&\textbf{0}\\
      n-b&mb&2nb&mb&n-b\\
      m&n+b&2&n+b&m\end{bmatrix}},\\ 
  K_{d-}&=K_d-K_{dt}\\ 
  &=a\cdot{\begin{bmatrix}-m&b-n&\textbf{0}&n-b&m\\
      -n-b&-mb&\textbf{0}&mb&n+b\\
      -2&-2nb &\textbf{0}&2nb&2\\
      -n-b&-mb&\textbf{0}&mb&n+b\\
      -m&b-n&\textbf{0}&n-b&m\end{bmatrix}},\\
\end{aligned}
\label{equ:sobel5_kd+-}
\end{equation}
which satisfy the symmetry requirements by calculating the sum and difference of
$K_{d}$ and $K_{dt}$.
If we efficiently use the two filters $K_{d+}$ and $K_{d-}$, then $G_{d}$
and $G_{dt}$ are easily obtained as follows:
\begin{equation}
\begin{aligned}
  G_d&=K_{d}\ast{I}=\frac{K_{d}^++K_{d}^-}{2}\ast{I}=\frac{G_{d}^++G_{d}^-}{2},\\ 
  G_{dt}&=K_{dt}\ast{I}=\frac{K_{d}^+-K_{d}^-}{2}\ast{I}=\frac{G_{d}^+-G_{d}^-}{2}.
\end{aligned}
\label{equ:sobel5_kd}
\end{equation}
For each row {\em u}, the convolution results $F_{ki}^u$ in the horizontal
direction using filter $K_{d+}$ can be obtained as follows:
\begin{equation}
  \begin{aligned}
    F_{k0}^u&=-am\cdot{p^0}+a(-n-b)\cdot{p^1}+(-2a)\cdot{p^2}\\
    &+a(-n-b)\cdot{p^3}+(-am)\cdot{p^4},\\
    F_{k1}^u&=a(b-n)\cdot{p^0}+(-amb)\cdot{p^1}+(-2anb)\cdot{p^2}\\
    &+(-amb)\cdot{p^3}+a(b-n)\cdot{p^4},\\
    F_{k2}^u&=0,\\
    F_{k3}^u&=a(n-b)\cdot{p^0}+(amb)\cdot{p^1}+(2anb)\cdot{p^2}\\
    &+(amb)\cdot{p^3}+a(n-b)\cdot{p^4},\\
    F_{k4}^u&=am\cdot{p^0}+a(n+b)\cdot{p^1}+(2a)\cdot{p^2}\\
    &+a(n+b)\cdot{p^3}+(am)\cdot{p^4}.
  \end{aligned}
  \label{equ:Fx_kd+}
\end{equation}
In addition, for a center row {\em v}, the results $G_{d+}^v$ can be obtained by
aggregating the four $F_{ki}$ from adjacent rows as follows:
\begin{equation}
  G_{d+}^v=F_{k0}^{v-2}+F_{k1}^{v-1}+F_{k3}^{v+1}+F_{k4}^{v+2}, v\geq{2}.
\label{equ:Gd+}
\end{equation}
Here, for the convenience of understanding, we use {\em ki} to represent the
vector index in filter $K_{d+}$, instead of using the thread index {\em i} in
Eq.~\ref{equ:Fx}.
Because the absolute values of the weights are symmetrical, for each row {\em
  u}, $F_{k3}^u$ and $F_{k4}^u$ are easily obtained using $F_{k1}^u$ and
$F_{k0}^u$:
\begin{equation}
  \begin{aligned}
    F_{k3}^u&=F_{-k1}^u=-F_{k1}^u,\\
    F_{k4}^u&=F_{-k0}^u=-F_{k0}^u,
  \end{aligned}
  \label{equ:Fx_34}
\end{equation}
and
\begin{equation}
  G_{d+}^v=F_{k0}^{v-2}+F_{k1}^{v-1}-F_{k1}^{v+1}-F_{k0}^{v+2}, v\geq{2}.
\label{equ:Gd+_new}
\end{equation}
Thus, we effectively reuse part of the intermediate results, as shown in 
Fig.~\ref{fig:kd+}, without repeating convolution operations for each row. 
Figure~\ref{fig:kd+}(a) shows the procedure of $G_{d+}$ and
Fig.~\ref{fig:kd+}(b) presents synchronous changes of on-chip register data. 
In Step 1, we use {\em k1} to convolve the second row instead of {\em k2}, because 
{\em k2} is a zero vector and does not affect the convolution result. This prepares the reused data required to ensure operation consistency
in each step.
Furthermore, we regard $F_{k3}^3$ and $F_{k4}^4$ as $F_{-k1}^3$ and $F_{-k0}^4$, 
respectively, to be able to discover the pattern of data reuse.
Then, $G_{d+}^2$ can be obtained according to Eq.~\ref{equ:Gd+_new} while 
loading the fifth row of the image.
After it succeeds, the sixth row begins to be loaded. Simultaneously, the
vectors from {\em k0} to {\em -k0} are strided down to convolve the rows centered on row 3. In Step 2, in addition to convolving the fifth row with 
{\em -k0}, only $F_{k0}^1$ and $F_{-k1}^4$ (green blocks) need to be recalculated 
because $F_{k1}^2$ can be reused. Compared with the original operations in Step 1, 
Step 2 significantly saves a quarter of the computation, ensuring that the filtering of $K_{d+}$ 
is efficiently performed.
Note that after Step 3, the second vectors {\em k1} used in each 
step are always the opposite of the previous step. However, this calculation can be 
reflected in the calculation of $G_{d+}$ without updating the register data.
This method can be repeatedly applied to the calculation of incremental rows, while the register index of the latest row is dynamically updated according to 
Eq.~\ref{equ:Fx_index_new}.

\begin{figure*}
\centering \includegraphics[width=6.7in]{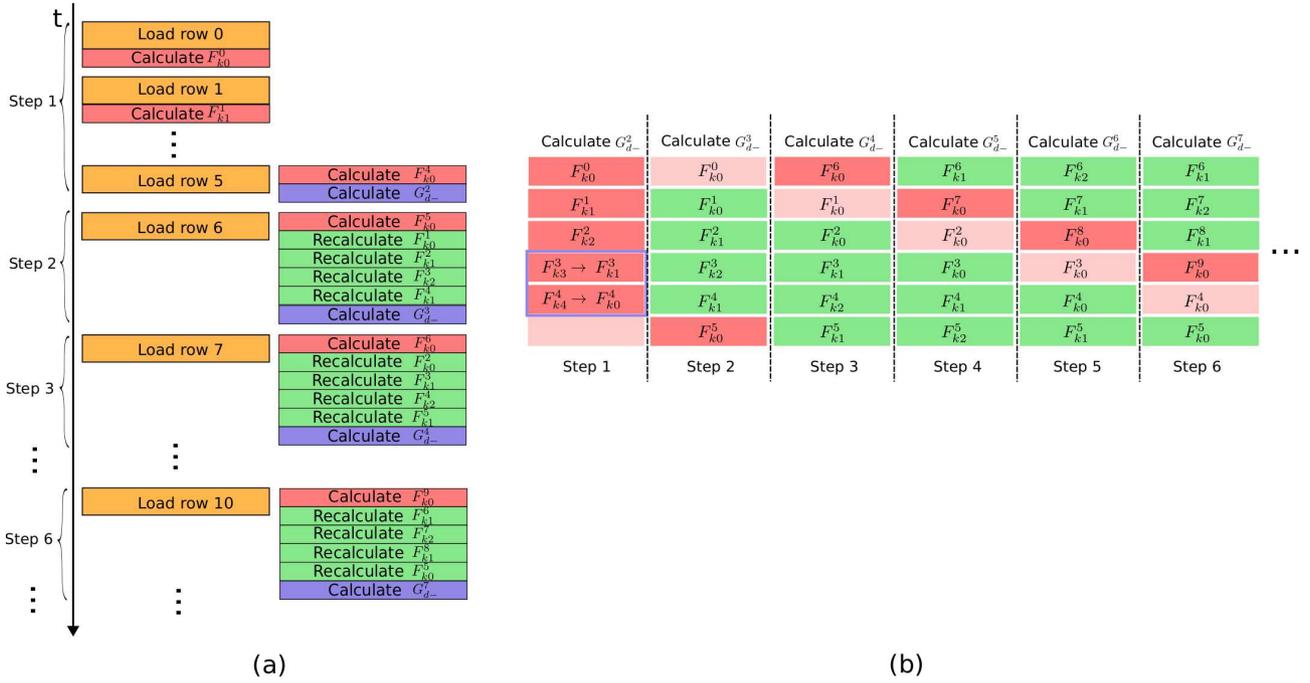}
\caption{Calculation for $G_{d-}$. (a) procedure. (b) Registers status.}
\label{fig:kd-}
\end{figure*}

For filter $K_{d-}$, the $F_{ki}^u$ in the horizontal direction can be
obtained as follows:
\begin{equation}
  \begin{aligned}
    F_{k0}^u&=-am\cdot{p^0}+a(b-n)\cdot{p^1}\\
    &+a(n-b)\cdot{p^3}+(am)\cdot{p^4},\\
    F_{k1}^u&=a(-n-b)\cdot{p^0}+(-amb)\cdot{p^1}\\
    &+(amb)\cdot{p^3}+a(n+b)\cdot{p^4},\\
    F_{k2}^u&=(-2a)\cdot{p^0}+(-2anb)\cdot{p^1}+2anb\cdot{p^3}+2a\cdot{p^4}\\
    F_{k3}^u&=F_{k1}^u\\
    F_{k4}^u&=F_{k0}^u.
  \end{aligned}
  \label{equ:Fd-}
\end{equation}
In addition, the $G_{d-}^v$ of a center row {\em v} can be obtained using
Eq.~\ref{equ:Gd-}:
\begin{equation}
  G_{d-}^v=F_{k0}^{v-2}+F_{k1}^{v-1}+F_{k2}^v+F_{k1}^{v+1}+F_{k0}^{v+2}, v\geq{2}.
\label{equ:Gd-}
\end{equation}
Figure~\ref{fig:kd-}(a) shows the procedure of $G_{d-}$, and
Fig.~\ref{fig:kd-}(b) presents the changes of on-chip register data
synchronously, which is the same method as $G_{d+}$. According to this 
figure, all convolutions for each row must be recalculated in each 
step because the $K_{d-}$ filter no longer has a zero vector in the 
horizontal direction, missing buffers that can be reused.
Inspired by Eq.~\ref{equ:sobel5_param}, we decompose $K_{d-}$ into the sum
of two products of two vectors as follows:
\begin{equation}
\begin{aligned}
  K_{d-}&=a\cdot{\begin{bmatrix}
      -m&b-n&\textbf{0}&n-b&m\\-n-b&-mb&\textbf{0}&mb&n+b\\-2&-2nb &\textbf{0}
      &2nb
      &2\\-n-b&-mb&\textbf{0}&mb&n+b\\-m&b-n&\textbf{0}&n-b&m\end{bmatrix}}\\ =&a\cdot\biggl(\begin{bmatrix}m\\n+b\\2\\n+b\\m\end{bmatrix}\times{\begin{bmatrix}-1&-b&\textbf{0}&b&1\end{bmatrix}}\\ &-\begin{bmatrix}mb+b-n\\nb+b^2-mb\\2b-2nb\\nb+b^2-mb\\mb-n+b\end{bmatrix}\times{\begin{bmatrix}0&-1&\textbf{0}&1&0\end{bmatrix}}\biggr).\\
\end{aligned}
\label{equ:sobel5_4}
\end{equation}
Equation~\ref{equ:sobel5_4} demonstrates that the first $1\times{5}$ horizontal
vector is the same as $K_{x}$, which means that its intermediate results can be
reused without recalculation. In addition, the second horizontal vector means
we only need to calculate the difference between columns 2 and 4.
Thus, the $G_{d-}^v$ of a center row {\em v} is easily obtained as follows:
\begin{equation}
  \begin{aligned}
    G_{d-}^v&=am\cdot{F^{f(v-2)}}+a(n+b)\cdot{F^{f(v-1)}}+2a\cdot{F^{f(v)}}\\
    &+a(n+b)\cdot{F^{f(v+1)}}+am\cdot{F^{f(v+2)}}\\
    &-a(mb+b-n)\cdot{D^{f(v-2)}}-a(nb+b^2-mb)\cdot{D^{f(v-1)}}\\
    &-a(2b-2nb)\cdot{D^{f(v)}}-a(nb+b^2-mb)\cdot{D^{f(v+1)}}\\
    &-a(mb-n+b)\cdot{D^{f(v+2)}},\\
  \end{aligned}
  \label{equ:Gd-_fast}
\end{equation}
and
\begin{equation}
    f(x)=x\bmod{6}, \hspace*{0.5cm} x\geq{0},\nonumber
\end{equation}
where $D$ denotes the convolution results under the second horizontal vector.
Therefore, although the on-chip registers must still be updated every
time, we only need to perform simple multiply-accumulate operations instead 
of multiple convolutions, significantly reducing the overall calculation amount 
and improving the processing speed. 
Thus far, the efficient calculation methods in all four directions for a
$5\times{5}$ Sobel operator have been introduced, and the final edge detection 
result can be obtained by integrating the respective results in these four 
directions according to Eq.~\ref{equ:sobel5_2}.

\begin{table*}
\caption{Speed performance of our four-directional Sobel operators. }
\begin{center}
\scalebox{0.845}{
\begin{tabular}{c|c|c|c|c|c|c|c|c|c|c|c|c} 
\toprule
\multicolumn{3}{c|}{}&\multicolumn{7}{c|}{\bf Execution time (\textmu s)}&\multicolumn{2}{c|}{\bf Throughput(GB/s)}\\
\midrule
{\bf Hardware} &{\bf Sobel operator} & {\bf Image size} &{\bf GM}  &{\bf SM}  &{\bf SM-P} &{\bf RG} &{\bf RG-v1} &{\bf RG-v2} &{\bf IO} &{\bf HToD} &{\bf DToH} &{\bf Speedup}\\         
\midrule
 \multirow{6}{*}{GTX1650Ti}&& $512\times{512}$
 &10.285&14.626&14.246&{\bf 6.766}&-&-&80.475&6.035&6.233&1.52\\
&$3\times{3}$&$1024\times{1024}$&43.792&55.639&54.455&{\bf 30.791}&-&-&316.52&6.115&6.287&1.42\\
&&$2048\times{2048}$&165.86&206.81&198.71&{\bf105.22}&-&-&1263.9&6.151&6.254&1.576\\
\cmidrule(lr){2-13}
& & $512\times{512}$ &29.712&30.913&28.606&24.905&20.884&{\bf 18.747}&80.475&6.035&6.223&1.585\\
&$5\times{5}$&$1024\times{1024}$&124.00&109.59&107.78&95.940&77.918&{\bf 66.225}&316.52&6.115&6.287&1.872\\
&&$2048\times{2048}$&424.37&418.50&411.91&350.36&286.01&{\bf 249.22}& 1263.9& 6.151  & 6.254  & 1.702  \\
\midrule
\multirow{6}{*}{Jetson AGX}&& $512\times{512}$ & 17.426  & 17.693  & 17.139  &{\bf  13.881}    &-&-& 17.049  & 26.512  & 34.488  & 1.255  \\
&$3\times{3}$&$1024\times{1024}$& 60.748  & 59.483  & 60.975  &{\bf  37.816  }&-&-& 65.737  & 28.802  & 32.440  & 1.606  \\
&&$2048\times{2048}$& 250.65  & 230.66  & 227.94  &{\bf  160.81  }&-&-& 914.23  & 10.504  & 7.501  & 1.559  \\
\cmidrule(lr){2-13}
& & $512\times{512}$ & 48.625  & 34.739  & 34.379  & 31.346  & 28.118  &{\bf  26.620  }& 17.049  & 26.512  & 34.488  & 1.827   \\
&$5\times{5}$&$1024\times{1024}$& 164.41  & 141.66  & 120.68  & 116.14  & 115.06  &{\bf  93.354  }& 65.737  & 28.802  & 32.440  & 1.761  \\
&&$2048\times{2048}$& 694.99  & 572.95  & 549.38  & 499.15  & 454.63  &{\bf  368.24  }& 914.23  & 10.504  & 7.501  & 1.887  \\
\bottomrule
\end{tabular}
}
\end{center}
\label{tbl:our_sobel}
\end{table*}


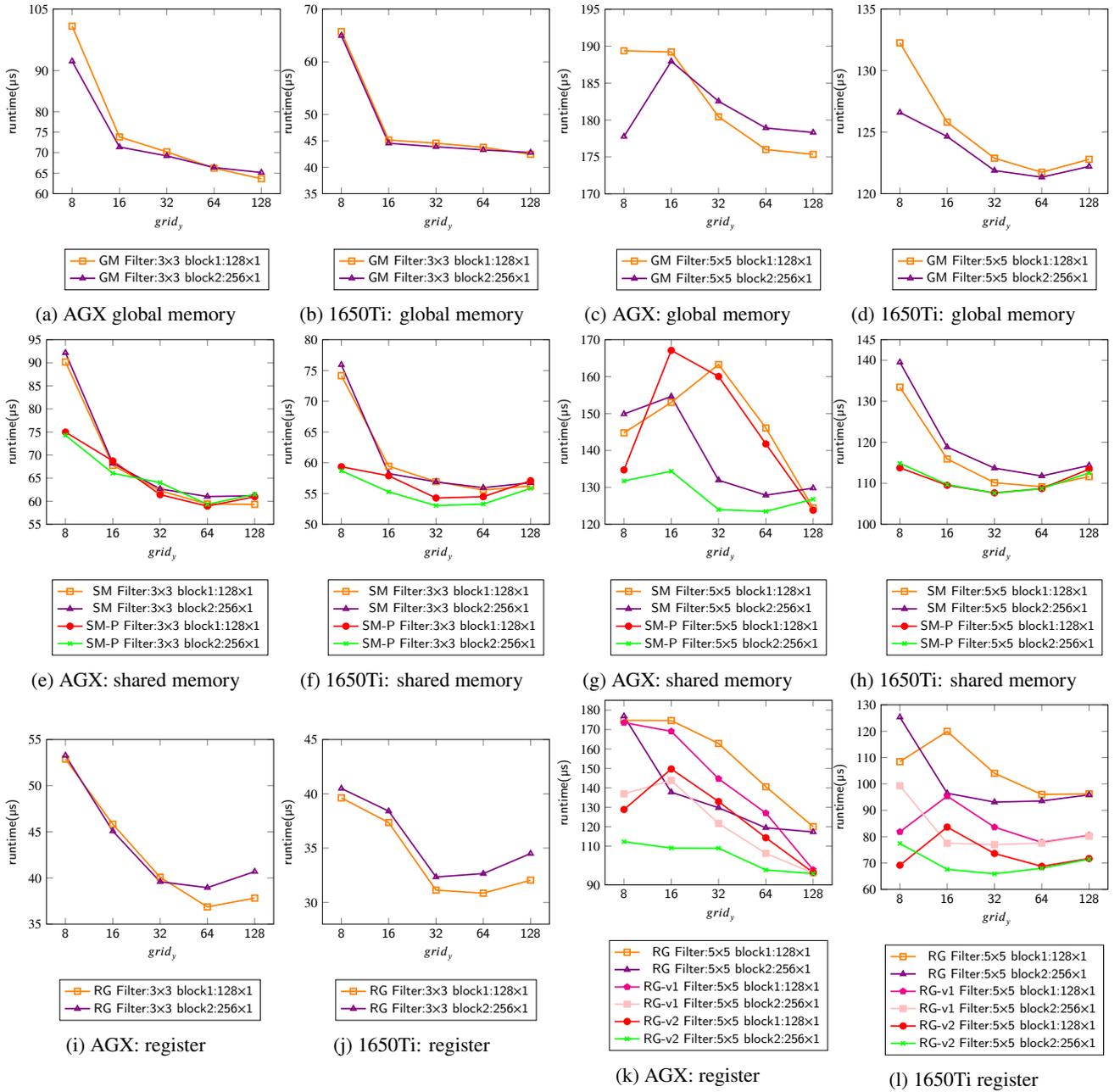
\begin{figure*}
\center
\begin{subfigure}{0.24\linewidth}
\begin{tikzpicture}[scale=0.65]
\begin{axis}[
    ylabel={runtime(\textmu s)}, 
    xlabel={\ $grid_y$},
    ymax=105,
    ymin=60,
    ytick={60,65,70,75,80,90,105},
    symbolic x coords={8,16,32,64,128}, 
    legend style={at={(0.5,-0.3)},anchor=north} 
    ]
\addplot[line width=1pt, mark=square,orange] plot coordinates 
{  
    (8,100.84)
    (16,73.826)
    (32,70.197)
    (64,66.227)
    (128,63.645)
};
\addlegendentry{GM Filter:3$\times$3 block1:128$\times$1}   
\addplot[line width=1pt,mark=triangle,violet] plot coordinates
{
    (8,92.284)
    (16,71.374)
    (32,69.207)
    (64,66.376)
    (128,65.158)
};
\addlegendentry{GM Filter:3$\times$3 block2:256$\times$1}
    \end{axis}
\end{tikzpicture}
\caption{AGX global memory}
\end{subfigure}
%
\begin{subfigure}{0.24\linewidth}
\begin{tikzpicture}[scale=0.65]
\begin{axis}[
    ylabel={runtime(\textmu s)}, 
    xlabel={\ $grid_y$},
    ymax=70,
    ymin=35,
    ytick={35,40,45,50,60,70},
    symbolic x coords={8,16,32,64,128}, 
    legend style={at={(0.5,-0.3)},anchor=north} 
    ]
\addplot[line width=1pt, mark=square,orange] plot coordinates 
{ 
    (8,65.730)
    (16,45.176)
    (32,44.568)
    (64,43.789)
    (128,42.464)
};
\addlegendentry{GM Filter:3$\times$3 block1:128$\times$1}   
\addplot[line width=1pt,mark=triangle,violet] plot coordinates
{
    (8,64.967)
    (16,44.554)
    (32,43.895)
    (64,43.296)
    (128,42.797)
};
\addlegendentry{GM Filter:3$\times$3 block2:256$\times$1}
\end{axis}
\end{tikzpicture}
\caption{1650Ti: global memory}
\end{subfigure}
%
\begin{subfigure}{0.24\linewidth}
\begin{tikzpicture}[scale=0.65]
\begin{axis}[
    ylabel={runtime(\textmu s)}, 
    xlabel={\ $grid_y$},
    ymax=195,
    ymin=170,
    ytick={170,175,180,185,190,195},
    symbolic x coords={8,16,32,64,128}, 
    legend style={at={(0.5,-0.3)},anchor=north} 
    ]
\addplot[line width=1pt, mark=square,orange] plot coordinates 
{ 
    (8,189.38)
    (16,189.22)
    (32,180.44)
    (64,176.00)
    (128,175.35)
};
\addlegendentry{GM Filter:5$\times$5 block1:128$\times$1}   
\addplot[line width=1pt,mark=triangle,violet] plot coordinates
{
    (8,177.76)
    (16,187.95)
    (32,182.54)
    (64,178.92)
    (128,178.31)
};
\addlegendentry{GM Filter:5$\times$5 block2:256$\times$1}
\end{axis}
\end{tikzpicture}
\caption{AGX: global memory}
\end{subfigure}
%
\begin{subfigure}{0.24\linewidth}
\begin{tikzpicture}[scale=0.65]
\begin{axis}[
    ylabel={runtime(\textmu s)}, 
    xlabel={\ $grid_y$},
    ymax=135,
    ymin=120,
    ytick={118,120,125,130,135},
    symbolic x coords={8,16,32,64,128}, 
    legend style={at={(0.5,-0.3)},anchor=north} 
    ]
\addplot[line width=1pt, mark=square,orange] plot coordinates 
{ 
    (8,132.26)
    (16,125.81)
    (32,122.88)
    (64,121.74)
    (128,122.79)
};
\addlegendentry{GM Filter:5$\times$5 block1:128$\times$1}   
\addplot[line width=1pt,mark=triangle,violet] plot coordinates
{
    (8,126.60)
    (16,124.64)
    (32,121.88)
    (64,121.34)
    (128,122.21)
};
\addlegendentry{GM Filter:5$\times$5 block2:256$\times$1}
\end{axis}
\end{tikzpicture}
\caption{1650Ti: global memory}
\end{subfigure}
%
\begin{subfigure}{0.24\linewidth}
\begin{tikzpicture}[scale=0.65]
\begin{axis}[
    ylabel={runtime(\textmu s)}, 
    xlabel={\ $grid_y$}, 
    ymax=95,
    ymin=55,
    ytick={55,60,65,70,75,80,85,90,95},
    symbolic x coords={8,16,32,64,128}, 
    legend style={at={(0.5,-0.3)},anchor=north} 
    ]
\addplot[line width=1pt, mark=square,orange] plot coordinates {
    (8,90.192)
    (16,67.764)
    (32,62.265)
    (64,59.406)
    (128,59.316)
};
\addlegendentry{SM Filter:3$\times$3 block1:128$\times$1}   
\addplot[line width=1pt, mark=triangle,violet] plot coordinates {
    (8,92.172)
    (16,68.219)
    (32,62.668)
    (64,61.02)
    (128,61.175)
};
\addlegendentry{SM Filter:3$\times$3 block2:256$\times$1}
\addplot[line width=1pt, mark=*,red] plot coordinates {
    (8,74.977)
    (16,68.735)
    (32,61.422)
    (64,58.939)
    (128,60.975)
};
\addlegendentry{SM-P Filter:3$\times$3 block1:128$\times$1}   
\addplot[line width=1pt, mark=x,green] plot coordinates {
    (8,74.286)
    (16,66.063)
    (32,64.028)
    (64,59.330)
    (128,61.503)
};
\addlegendentry{SM-P Filter:3$\times$3 block2:256$\times$1}
\end{axis}
\end{tikzpicture}
\caption{AGX: shared memory}
\end{subfigure}
%
\begin{subfigure}{0.24\linewidth}
\begin{tikzpicture}[scale=0.65]
\begin{axis}[
    ylabel={runtime(\textmu s)}, 
    xlabel={\ $grid_y$},
    ymax=80,
    ymin=50,
    ytick={50,55,60,65,70,75,80},
    symbolic x coords={8,16,32,64,128}, 
    legend style={at={(0.5,-0.3)},anchor=north} 
    ]
\addplot[line width=1pt, mark=square,orange] plot coordinates 
{ 
    (8,74.160)
    (16,59.441)
    (32,56.930)
    (64,55.572)
    (128,56.222)
};
\addlegendentry{SM Filter:3$\times$3 block1:128$\times$1}   
\addplot[line width=1pt,mark=triangle,violet] plot coordinates
{
    (8,75.916)
    (16,58.251)
    (32,56.886)
    (64,55.953)
    (128,56.747)
};
\addlegendentry{SM Filter:3$\times$3 block2:256$\times$1}
\addplot[line width=1pt, mark=*,red] plot coordinates {
    (8,59.349)
    (16,57.879)
    (32,54.274)
    (64,54.486)
    (128,57.076)
};
\addlegendentry{SM-P Filter:3$\times$3 block1:128$\times$1}   
\addplot[line width=1pt, mark=x,green] plot coordinates {
    (8,58.696)
    (16,55.278)
    (32,53.035)
    (64,53.295)
    (128,55.826)
};
\addlegendentry{SM-P Filter:3$\times$3 block2:256$\times$1}
\end{axis}
\end{tikzpicture}
\caption{1650Ti: shared memory}
\end{subfigure}
%
\begin{subfigure}{0.24\linewidth}
\begin{tikzpicture}[scale=0.65]
\begin{axis}[
    ylabel={runtime(\textmu s)}, 
    xlabel={\ $grid_y$}, 
    ymax=170,
    ymin=120,
    ytick={120,130,140,150,160,170},
    symbolic x coords={8,16,32,64,128}, 
    legend style={at={(0.5,-0.3)},anchor=north} 
    ]
\addplot[line width=1pt, mark=square,orange] plot coordinates {
    (8,144.77)
    (16,153.01)
    (32,163.28)
    (64,146.08)
    (128,124.45)
};
\addlegendentry{SM Filter:5$\times$5 block1:128$\times$1}   
\addplot[line width=1pt, mark=triangle,violet] plot coordinates {
    (8,149.87)
    (16,154.63)
    (32,131.94)
    (64,127.88)
    (128,129.80)
};
\addlegendentry{SM Filter:5$\times$5 block2:256$\times$1}
\addplot[line width=1pt, mark=*,red] plot coordinates {
    (8,134.72)
    (16,167.12)
    (32,160.03)
    (64,141.76)
    (128,123.80)
};
\addlegendentry{SM-P Filter:5$\times$5 block1:128$\times$1}   
\addplot[line width=1pt, mark=x,green] plot coordinates {
    (8,131.75)
    (16,134.37)
    (32,123.99)
    (64,123.49)
    (128,126.80)
};
\addlegendentry{SM-P Filter:5$\times$5 block2:256$\times$1}
\end{axis}
\end{tikzpicture}
\caption{AGX: shared memory}
\end{subfigure}
%
\begin{subfigure}{0.24\linewidth}
\begin{tikzpicture}[scale=0.65]
\begin{axis}[
    ylabel={runtime(\textmu s)}, 
    xlabel={\ $grid_y$},
    ymax = 145,
    ymin = 100,
    ytick={100,110,120,130,140,145},
    symbolic x coords={8,16,32,64,128}, 
    legend style={at={(0.5,-0.3)},anchor=north} 
    ]
\addplot[line width=1pt, mark=square,orange] plot coordinates 
{ 
    (8,133.42)
    (16,115.92)
    (32,110.14)
    (64,109.15)
    (128,111.63)
};
\addlegendentry{SM Filter:5$\times$5 block1:128$\times$1}   
\addplot[line width=1pt,mark=triangle,violet] plot coordinates
{
    (8,139.50)
    (16,118.83)
    (32,113.69)
    (64,111.78)
    (128,114.34)
};
\addlegendentry{SM Filter:5$\times$5 block2:256$\times$1}
\addplot[line width=1pt, mark=*,red] plot coordinates {
    (8,113.77)
    (16,109.51)
    (32,107.66)
    (64,108.70)
    (128,113.43)
};
\addlegendentry{SM-P Filter:5$\times$5 block1:128$\times$1}   
\addplot[line width=1pt, mark=x,green] plot coordinates {
    (8,114.86)
    (16,109.70)
    (32,107.60)
    (64,108.71)
    (128,112.51)
};
\addlegendentry{SM-P Filter:5$\times$5 block2:256$\times$1}
\end{axis}
\end{tikzpicture}
\caption{1650Ti: shared memory}
\end{subfigure}
%
\begin{subfigure}{0.24\linewidth}
\begin{tikzpicture}[scale=0.65]
\begin{axis}[
    ylabel={runtime(\textmu s)}, 
    xlabel={\ $grid_y$},
    ymax=55,
    ymin=35,
    ytick={35,40,45,50,55},
    symbolic x coords={8,16,32,64,128}, 
    legend style={at={(0.5,-0.3)},anchor=north} 
    ]
\addplot[line width=1pt, mark=square,orange] plot coordinates { 
    (8,52.877)
    (16,45.823)
    (32,40.082)
    (64,36.863)
    (128,37.816)
};
\addlegendentry{RG Filter:3$\times$3 block1:128$\times$1}   
\addplot[line width=1pt, mark=triangle,violet] plot coordinates {
    (8,53.274)
    (16,45.074)
    (32,39.588)
    (64,38.949)
    (128,40.689)
};
\addlegendentry{RG Filter:3$\times$3 block2:256$\times$1}
\end{axis}
\end{tikzpicture}
\caption{AGX: register}
\end{subfigure}
%
\begin{subfigure}{0.24\linewidth}
\begin{tikzpicture}[scale=0.65]
\begin{axis}[
    ylabel={runtime(\textmu s)}, 
    xlabel={\ $grid_y$},
    ymax=45,
    ymin=28,
    ytick={30,35,40,45},
    symbolic x coords={8,16,32,64,128}, 
    legend style={at={(0.5,-0.3)},anchor=north} 
    ]
\addplot[line width=1pt, mark=square,orange] plot coordinates 
{ 
    (8,39.626)
    (16,37.355)
    (32,31.126)
    (64,30.846)
    (128,32.045)
};
\addlegendentry{RG Filter:3$\times$3 block1:128$\times$1}   
\addplot[line width=1pt,mark=triangle,violet] plot coordinates
{
    (8,40.494)
    (16,38.406)
    (32,32.339)
    (64,32.652)
    (128,34.501)
};
\addlegendentry{RG Filter:3$\times$3 block2:256$\times$1}
\end{axis}
\end{tikzpicture}
\caption{1650Ti: register}
\end{subfigure}
%
\begin{subfigure}{0.24\linewidth}
\begin{tikzpicture}[scale=0.65]
\begin{axis}[
    ylabel={runtime(\textmu s)}, 
    xlabel={\ $grid_y$},
    ymax=185,
    ymin=90,
    ytick={90,110,120,130,140,150,160,170,180},
    symbolic x coords={8,16,32,64,128}, 
    legend style={at={(0.5,-0.3)},anchor=north} 
    ]
\addplot[line width=1pt, mark=square,orange] plot coordinates { 
    (8,174.66)
    (16,174.59)
    (32,162.81)
    (64,140.51)
    (128,120.00)
};
\addlegendentry{RG Filter:5$\times$5 block1:128$\times$1}   
\addplot[line width=1pt, mark=triangle,violet] plot coordinates {
    (8,176.86)
    (16,137.73)
    (32,129.73)
    (64,119.40)
    (128,117.25)
};
\addlegendentry{RG Filter:5$\times$5 block2:256$\times$1}
\addplot[color=magenta,line width=1pt, mark=pentagon*] plot coordinates { 
    (8,173.52)
    (16,169.05)
    (32,144.64)
    (64,126.99)
    (128,97.770)
};
\addlegendentry{RG-v1 Filter:5$\times$5 block1:128$\times$1}   
\addplot[color=pink,line width=1pt, mark=square*] plot coordinates {
    (8,136.89)
    (16,143.77)
    (32,121.65)
    (64,106.24)
    (128,96.062)
};
\addlegendentry{RG-v1 Filter:5$\times$5 block2:256$\times$1}
\addplot[line width=1pt, mark=*,red] plot coordinates { 
    (8,128.80)
    (16,149.67)
    (32,132.93)
    (64,114.26)
    (128,96.374)
};
\addlegendentry{RG-v2 Filter:5$\times$5 block1:128$\times$1}   
\addplot[line width=1pt, mark=x,green] plot coordinates {
    (8,112.26)
    (16,108.96)
    (32,108.88)
    (64,97.693)
    (128,95.799)
};
\addlegendentry{RG-v2 Filter:5$\times$5 block2:256$\times$1}
\end{axis}
\end{tikzpicture}
\caption{AGX: register}
\end{subfigure}
%
\begin{subfigure}{0.24\linewidth}
\begin{tikzpicture}[scale=0.65]
\begin{axis}[
    ylabel={runtime(\textmu s)}, 
    xlabel={\ $grid_y$},
    ymax=130,
    ymin=60,
    ytick={60,70,80,90,100,110,120,130},
    symbolic x coords={8,16,32,64,128}, 
    legend style={at={(0.5,-0.3)},anchor=north} 
    ]
\addplot[line width=1pt, mark=square,orange] plot coordinates 
{ 
    (8,108.45)
    (16,119.92)
    (32,104.04)
    (64,96.034)
    (128,96.262)
};
\addlegendentry{RG Filter:5$\times$5 block1:128$\times$1}   
\addplot[line width=1pt,mark=triangle,violet] plot coordinates
{
    (8,125.31)
    (16,96.492)
    (32,93.126)
    (64,93.548)
    (128,95.878)
};
\addlegendentry{RG Filter:5$\times$5 block2:256$\times$1}
\addplot[color=magenta,line width=1pt, mark=pentagon*] plot coordinates { 
    (8,81.847)
    (16,95.389)
    (32,83.620)
    (64,77.832)
    (128,80.498)
};
\addlegendentry{RG-v1 Filter:5$\times$5 block1:128$\times$1}  
\addplot[color=pink,line width=1pt, mark=square*] plot coordinates {
    (8,99.309)
    (16,77.501)
    (32,77.008)
    (64,77.556)
    (128,80.221)
};
\addlegendentry{RG-v1 Filter:5$\times$5 block2:256$\times$1}
\addplot[line width=1pt, mark=*,red] plot coordinates { 
    (8,69.157)
    (16,83.646)
    (32,73.610)
    (64,68.755)
    (128,71.687)
};
\addlegendentry{RG-v2 Filter:5$\times$5 block1:128$\times$1}  
\addplot[line width=1pt, mark=x,green] plot coordinates {
    (8,77.406)
    (16,67.599)
    (32,65.909)
    (64,68.016)
    (128,71.469)
};
\addlegendentry{RG-v2 Filter:5$\times$5 block2:256$\times$1}
\end{axis}
\end{tikzpicture}
\caption{1650Ti register}
\end{subfigure}
\caption{Speed comparison of four-directional Sobel operators for a $1024\times{1024}$ image in different
  resource configurations.}
\label{fig:blockset}
\end{figure*}

\begin{table*}
\caption{Speed comparisons of two-directional Sobel operators with other methods. }
\begin{center}
\begin{tabular}{l|c|c|c|l|c|c}
\toprule
{\bf Method}& {\bf Operator size} & {\bf Image size} &{\bf Runtime$^{*}$ (ms)} &{\bf Hardware} &{\bf MPS}\ &{\bf MPS/C}\\
\midrule
\multirow{4}{*}{\bf SobelGPU-Jetson}& $3\times{3}$&$1024\times{1024}$&{\bf 0.074}&\multirow{4}{*}{Jetson AGX}&{\bf 1.41E10}&{\bf 2.77E7}\\
& &$2048\times{2048}$&{\bf 0.519}&&{\bf 8.07E9}&{\bf 1.58E7}\\
& $5\times{5}$&$1024\times{1024}$&{\bf 0.085}&&{\bf 1.24E10}&{\bf 2.42E7}\\
& &$2048\times{2048}$&{\bf 0.552}&&{\bf 7.59E9}&{\bf 1.48E7}\\
\midrule
\multirow{4}{*}{\bf SobelGPU-GTX}& $3\times{3}$&$1024\times{1024}$&0.190&\multirow{4}{*}{GTX 1650Ti}&5.51E9&5.38E6\\
& &$2048\times{2048}$&0.740&&5.66E9&5.53E6\\ 
& $5\times{5}$&$1024\times{1024}$&0.199&&5.27E9&5.14E6\\
& &$2048\times{2048}$&0.763&&5.49E9&5.36E6\\
\midrule
\multirow{4}{*}{OpenCV-GPU 1}& $3\times{3}$&$1024\times{1024}$&0.512&\multirow{4}{*}{Jetson AGX}&2.05E9&4E6\\
& &$2048\times{2048}$&1.778&&2.36E9&4.6E6\\
& $5\times{5}$&$1024\times{1024}$&0.566&&1.85E9&3.62E6\\
& &$2048\times{2048}$&1.832&&2.29E9&4.47E6\\
\midrule
\multirow{4}{*}{OpenCV-GPU 2}& $3\times{3}$&$1024\times{1024}$&2.43&\multirow{4}{*}{GTX 1650Ti}&4.31E8&4.21E5\\
& &$2048\times{2048}$&9.82&&4.27E8&4.18E5\\
& $5\times{5}$&$1024\times{1024}$&2.53&&4.14E8&4.05E5\\
& &$2048\times{2048}$&9.90&&4.24E8&4.14E5\\
\midrule
Xiao~\cite{citation11}& $3\times{3}$&$1024\times{1024}$&5.48$^{\dag}$&GTX 1070&1.91E8&9.97E4\\
&&$2048\times{2048}$&18.95$^{\dag}$&&2.21E8&1.15E5\\
\midrule
Zahra~\cite{citation20}& $3\times{3}$&$512\times{512}$&3.62&GTX 550Ti&7.24E7&3.77E5\\
&&$1024\times{1024}$&14.74&&7.11E7&3.7E5\\
\midrule
\multirow{4}{*}{Theodora~\cite{citation21}}& $3\times{3}$&$1024\times{1024}$&0.601&\multirow{4}{*}{GTX 1060}&1.74E9&1.36E6\\
& &$2048\times{2048}$& 0.926&& 4.52E9&3.53E6\\
&$5\times{5}$&$1024\times{1024}$&0.837&&1.25E9&9.79E5\\
& &$2048\times{2048}$&1.174&&3.57E9&2.79E6\\
\midrule
Dore~\cite{citation22}& $3\times{3}$&$1024\times{1024}$&11.01$^{\dag}$&GTX 470&9.5E7&2.1E5\\
&&$2048\times{2048}$&84.023$^{\dag}$&&5E7&1.1E5\\
\midrule
You~\cite{citation23}& $3\times{3}$&$1024\times{768}$&5&DE1-SoC&1.57E8&-\\
&&$1920\times{1080}$&15&&1.38E8&-\\
\midrule
Sato~\cite{citation24}& $3\times{3}$&$512\times{512}$&1.1&Cyclone II&2.38E8&-\\
&&$1024\times{1024}$&4.37&&2.39E8&-\\
\midrule
Tim~\cite{citation25}& $3\times{3}$&$1280\times{1024}$&31&ZYNQ 7030&4.22E7&-\\
\bottomrule
\end{tabular}
\end{center}
\begin{tablenotes}
    \small
     \item[1] \hspace*{1.5cm}*: Runtime includes the kernel execution time and data loading time. $^{\dag}$: Only the kernel execution time is included.
   \end{tablenotes}
\label{tbl:comparison}
\end{table*}

\section{Evaluation and Discussion}
\subsection{Evaluation Platforms}
We implemented our multi-directional 5$\times$5 Sobel operator kernel on an embedded Jetson AGX Xavier GPU and Nvidia GTX 1650Ti mobile GPU, because as widely used mobile GPUs, they have recently been used in some studies to handle systems that combine Sobel operators with other upper-layer applications~\cite{citation17}~\cite{citation18}~\cite{citation19}. At the same time, these two kinds of GPUs have different architectures, which determine that the same CUDA kernel often reflects utterly different performance.  
Jetson AGX Xavier is a powerful platform, built on an Nvidia Volta GPU with 512 cores and 
shares physical memory with the center processor. Users can allow both the host and device 
to access the shared data by applying for managed memory, reducing the impact of data transmission.
In contrast, the GTX 1650Ti is built on a Turing architecture with 1024 cores and has dedicated 
memory. Although it has a higher calculation capability than Jetson AGX Xavier, the data 
it processes must be first transferred from the host memory to the device memory 
through the PCIe bus, which increases the burden of IO.
Evaluating our kernel on both platforms helps us fully understand its performance, and 
users can choose different solutions according to their requirements. 

\subsection{Evaluation of Kernel Performance}
We evaluated the performance using three
different sizes of images: 512$\times{512}$, $1024\times1024$, and
$2048\times{2048}$. To evaluate kernel performance more
comprehensively, we used global and shared memory as storage mediums other than the register.
Additionally, we compared the 3$\times$3 operator.
Table~\ref{tbl:our_sobel} lists the speed performance of our four-directional
kernels. {\em GM}, {\em SM}, and {\em RG} represent the original methods using
global memory, shared memory, and registers, respectively.
{\em SM-P} represents the method that covers transmission latency by adding
the prefetching mechanism to {\em SM}.
{\em RG-v1} indicates that we transformed the original diagonal filters $K_{d}$ and
$K_{dt}$ into $K_{d+}$ and $K_{d-}$; and {\em RG-v2} the method that
further decomposes $K_{d-}$ based on {\em RG-v1}.
All {\em RG} series kernels are equipped with the prefetching mechanism.
Because the calculation of the 3$\times$3 operator in the diagonal directions is not
complicated, we only perform {\em RG-v1} and {\em RG-v2} to the 5$\times$5
operator.
{\em Speedup} denotes the ratio of {\em GM} to {\em RG} at runtime. 
For each kernel, we used the {\em NVprof} profiling tool to measure the kernel execution time 100 times and took the average value as the final execution time. Also, we calculated the standard deviation of the execution time of each kernel, ranging from 0.06 to 5.05, which fully proves the robustness of our measurement results.
Regardless of the platform, our kernel achieved a 1.3x speedup, and 
the maximum even reached over 1.8x.
For the degraded case, the {\em SM} series kernels on a GTX 1650Ti are slower than those of {\em GM}.
This is because using the shared memory without optimization only increases data 
transmission costs and reduces kernel performance.
Although using the prefetching mechanism can hide latency and reduce
the execution time by 19{\textmu s} on average compared with {\em GM}, the performance of {\em SM-P} for a 3$\times$3 operator on GTX 1650Ti is still lower than that of {\em GM}. This implies that, for a 3$\times$3 operator, using shared memory as the
storage medium is not a good choice.
The execution time of the kernels increases linearly with the image size on both platforms: 
approximately 4x on both platforms. This steady
change indirectly indicates that our method can fully utilize hardware
resources.
Additionally, in almost all cases, the introductions of {\em SM-P}, {\em RG-v1} 
and {\em RG-v2} gradually introduce a reduction in execution time, indicating that 
our proposed methods are effective and have high robustness. 
Particularly for the 5$\times$5 operator, the speed of our accelerated kernel {\em RG-v2} (66.225\textmu s) 
is only 33\% slower than the original 3$\times$3 kernel {\em GM} (43.792\textmu s), enabling us to use the 5$\times$5 Sobel 
operator with higher detection precision instead of 3$\times$3 in the future.
{\em IO} denotes the data transmission time required according to the hardware architecture. 
As mentioned, because Jetson AGX Xavier GPU shares the physical memory between the GPU and CPU, 
it does not cost too much on IO. By contrast, the IO costs required on GTX 1650Ti are much 
higher than the kernel execution. The {\em Throughput} metrics between the host and the device in both directions also reflect the same issue. The throughputs we achieved on Jetson AGX Xavier are much higher than those of GTX 1650Ti, but still far from theoretical values. This means that our kernel is memory limited and could be further ameliorated by processing larger images or video streams.
Because our kernel is implemented in CUDA, the block size configuration is closely
related to its performance. Therefore, we provide different combinations of
block configurations to the kernels and perform the 3$\times$3 and
5$\times$5 four-directional Sobel operators on the 1024$\times$1024 image shown in Fig.~\ref{fig:blockset}.
Each row of graphs represents different storage mediums used in our kernels.
The graphs in columns 1 and 2 show the processing results of the
3$\times$3 operator under different block configurations and platforms,
and columns 3 and 4 show those for 5$\times$5.
In each sub-graph, the x-axis represents the number of {\em grid.y}, which is
determined by the number of image sizes and threads allocated in the y direction
within each block;
the larger the number, the fewer rows each block processes in the y-axis
direction.
The y-axis represents the execution time of these kernels.
The methods performed here are the same as those tested in Table~\ref{tbl:our_sobel}. 
The difference is that we specified {\em block1:
  (128,1)} and {\em block2: (256,1)} configurations for each method.
According to the results, {\em block1} and {\em block2} do not affect much 
under the same method in most cases. 
Moreover, the speed relationship between these methods is the same as that shown 
in Table~\ref{tbl:our_sobel}. Therefore, we predict that the individual difference 
is caused by changes in thread parallelism resulting from different register usage 
rates.
For each method, a large {\em grid.y} typically shows a high performance because a 
higher number of blocks indicates a more significant number of active blocks in 
parallel. Thus, the kernel can use hardware resources better, resulting in better 
performance.
Table~\ref{tbl:comparison} shows the speed comparison of our kernel with 
other methods in two directions. The comparison objects are fast Sobel 
operators published in recent years, and each study provides their operator 
sizes, image sizes, execution time, and required hardware platforms. 
Here, {\em Runtime} includes the kernel execution time 
and data loading time from the host memory to the device.
The hardware used by these studies is primarily divided into two categories: GPUs
and field programmable gate arrays (FPGAs). Both are the most widely used algorithm accelerators today. To compare the computing capability of these operators based on 
different hardware, we list the mega-pixel per second (MPS) values of all these studies. Additionally, 
we use the mega-pixel per second per core (MPS/C) parameter, which represents the number 
of pixels processed per second by each core, to normalize their processing capabilities on different GPUs.
According to the {\em Runtime}, our kernels based on AGX are faster than those based on 1650Ti, contrary to the results shown in Table~\ref{tbl:comparison}. This is because of the considerable time required by the IO, resulting in a decrease in overall throughput. 
Compared with other studies, our operators are much faster in each case. Particularly for OpenCV-GPU, the most commonly used method 
in image edge detection, the processing speed is approximately 3.3x to 13.3x slower than 
our kernels. This is because the OpenCV-GPU treads the Sobel operator as a 2D convolution filter by default, and ours is actually further optimized on the basis of two 1D separable kernels. Besides, the OpenCV-GPU does not provide the functions in the diagonal directions, where our kernel has a considerable advantage. Xiao~\cite{citation11}, Zahra~\cite{citation20}, and Dore~\cite{citation22} implemented their 
fast 3$\times$3 Sobel operators on GPUs, and You~\cite{citation23}, 
Sata~\cite{citation24}, and Tim~\cite{citation25} implemented on FPGAs. They all achieved 
real-time processing on a large-scale image, but remain at the milliseconds level, 
leaving little processing time for upper-layer applications.
Theodora~\cite{citation21} implemented a complete version evaluation, including the 
combination of two Sobel operators and two images of different sizes. Their
execution times are approximately 1.3x to 4.2x longer than ours, even using a GTX 1060 GPU 
superior to our GTX 1650Ti.
According to the {\em MPS} and {\em MPS/C}, our numbers exceed those of other studies, 
demonstrating that our kernels have an overwhelming advantage.
\begin{figure}
\centering
\includegraphics[width=3in]{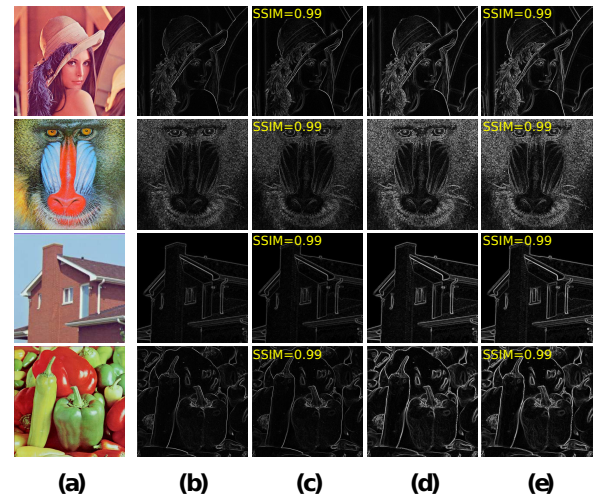}
\caption{Confirmation of edge detection results using 5x5 Sobel operators. (a) Two-directional: OpenCV-GPU kernel. (b) Two-directional: Our RG kernel. (c) Four-directional: Our GM kernel. (d) Four-directional: Our RG-v2 kernel.}
\label{fig:out_correct}
\end{figure}

 To confirm the correctness, we listed edge detection results of four images shown in Fig.~\ref{fig:out_correct}. For each image, we perform the edge detection using four different GPU kernels, including the two-directional OpenCV-GPU kernel (Fig.~\ref{fig:out_correct}(b)); our two-directional RG kernel (Fig.~\ref{fig:out_correct}(c)); our four-directional GM kernel (Fig.~\ref{fig:out_correct}(d)) and our four-directional RG-v2 kernel (Fig.~\ref{fig:out_correct}(e)). All the sizes of kernels are 5$\times$5. Here, because kernels (b) and (d) are implemented by the most primitive method, we take them as reference objects, and calculate the Structure Similarity Index Measure (SSIM) values of (c) and (e) relative to them respectively. SSIM is an indicator to measure the similarity of two images, and is calculated from the three image features of luminance, contrast and structure. It can be calculated as follow:
\begin{equation}
    \begin{aligned}
 SSIM(x, y) &= \frac{(2\mu_x\mu_{y} + C_1)(2\sigma_{xy}+C_2)}{(\mu_{x}^{2}+\mu_{y}^{2}+C_1)(\sigma_{x}^{2}+\sigma_{y}^{2}+C_2)}.      
    \end{aligned}
\label{SSIM LOSS}
\end{equation}
Here, $\mu$ and $\sigma^2$ represent the mean value and variance of the image,
respectively, and $\sigma_{xy}$ represents the covariance between the image x and image y. Finally, $C_1$ and $C_2$ are two constants used to avoid instability when $\mu_{x}^{2}+\mu_{y}^{2}$ or $\sigma_{x}^{2}+\sigma_{y}^{2}$
are close to zero.
The closer SSIM to 1 indicates, the higher the similarity of two images. According to the high SSIM values of 0.99 shown in Fig.~\ref{fig:out_correct}, it can be observed that the proposed acceleration method can guarantee the correctness of Sobel operators in both two and four directions.

\section{Conclusion}
This paper proposed a fast GPU kernel for a four-directional
$5\times{5}$ Sobel operator that entirely uses the register resource.
We improved the Sobel operator from two perspectives: computer architecture 
and mathematics. In computer architecture, we focused on fully using registers 
with the help of warp-level primitives without utilizing global memory and shared 
memory. Simultaneously, we introduced the prefetching mechanism to hide the system 
latency caused by data transmission. Concerning mathematics, we proposed a two-step 
optimization method for the Sobel operator with complex patterns in diagonal 
directions, enabling us to fully reuse intermediate results and significantly 
improve the execution efficiency of the kernel. Extensive experiments prove that our kernel has high robustness, with significant improvements 
in detection speed for images of different sizes. Furthermore, our kernel achieves 
6.7x and 13x improvements in processing speed compared with the OpenCV-GPU library on two different GPUs. To the best of our 
knowledge, the proposed kernel is currently the fastest kernel based on GPUs.
To further facilitate our kernel's application, we plan to combine it with high-level applications such as object detection. In addition, for the bottleneck problem of IO, 
we predict that stream processing can efficiently reduce the latency caused by data 
transmission. In addition, the burden of on-chip computation not being too heavy must be ensured. These concerns will be addressed in future work.

\section*{CRediT Author Statement}
{\bf Qiong Chang}: algorithm design, basic code writing, experiment design, draft manuscript writing. 
{\bf Xiang Li}: code optimization, experiment, manuscript co-writing.
{\bf Yun Li}: supervision, experiment environment preparation, writing-reviewing.
{\bf Jun Miyazaki}: supervision, writing-reviewing and editing.
All authors contributed to discussions. 
Qiong Chang and Xiang Li contributed equally to this work.

\section*{Conflict of interest}
The authors declare that they have no conflict of interest.

\end{document}